\documentclass[showpacs,twocolumn,floats]{revtex4}
\usepackage{graphicx}
\usepackage{amsmath}
\begin{document}
\title{Quasiclassical fluctuations of the superconductor proximity gap in a
  chaotic system}
\author{M.C. Goorden, Ph.\ Jacquod\footnote{Present address: Department of
  Theoretical Physics, University of Geneva, 24 Quai Ernest Ansermet, CH-1211
  Geneva 4, Switzerland.}, and C.W.J. Beenakker}
\affiliation{Instituut-Lorentz, Universiteit Leiden, P.O. Box 9506,
2300 RA Leiden, The Netherlands}
\date{30 June 2003}
\begin{abstract}
We calculate the sample-to-sample fluctuations in the excitation gap of a
chaotic dynamical system coupled by a narrow lead to a superconductor. Quantum
fluctuations on the order of magnitude of the level spacing, predicted by
random-matrix theory, apply if $\tau_E\ll\hbar/E_T$ (with $\tau_E$ the
Ehrenfest time and $E_T$ the Thouless energy). For $\tau_E\agt\hbar/ E_T$ the
fluctuations are much greater than the level spacing. We demonstrate the
quasiclassical nature of the gap fluctuations in the large-$\tau_E$ regime by
correlating them to an integral over the classical dwell-time distribution.
\end{abstract}
\pacs{74.45.+c, 03.65.Sq, 05.45.Mt, 74.78.Na}
\maketitle

The universality of statistical fluctuations is one of the most profound
manifestations of quantum mechanics in mesoscopic systems
\cite{Imr97}. Classically, the conductance $g$ of a disordered metal (measured
in the fundamental unit $2e^2/h$) would fluctuate from sample to sample by an
amount of order $(l/L)^{3/2}\ll 1$, with $l$ the mean free path and $L$ the
length of the conductor \cite{Gal91}.  Quantum mechanical interference
increases the fluctuations to order unity, independent of disorder or sample
length. This is the phenomenon of universal conductance fluctuations
\cite{Alt85,Lee85}. The same universality applies to a variety of other
properties of disordered metals and superconductors, and random-matrix theory
(RMT) provides a unified description \cite{Bee93}.

Chaotic systems (for example, a quantum dot in the shape of a stadium) share
much of the phenomenology of disordered systems: The same universality of
sample-to-sample fluctuations exists \cite{Bee97,Guh98,Alh00}. What is
different is the appearance of a new time scale, below which RMT breaks down
\cite{Lod98,Ale96}. This time scale is the Ehrenfest time $\tau_E$, which
measures how long it takes for a wave packet of minimal size to expand over
the entire available phase space. If $\tau_E$ is larger than the mean dwell
time $\tau_D$ in the system (the reciprocal of the Thouless energy
$E_T=\hbar/2\tau_D$), then interference effects are  inoperative. A chaotic
system with conductance $g\times 2e^2/h$, level spacing $\delta$, and Lyapunov
exponent $\lambda$ has $\tau_D=2\pi\hbar/g\delta$ and $\tau_E=
\lambda^{-1}\ln{(g\tau_0/\tau_D)}$, with $\tau_0$ the time of flight across
the system \cite{Vav03}. The defining characteristic of the Ehrenfest time is
that it scales logarithmically with $\hbar$, or equivalently, logarithmically
with the system size over Fermi wavelength \cite{Zas81}.

The purpose of this paper is to investigate what happens to mesoscopic
fluctuations if the Ehrenfest time becomes comparable to, or larger than, the
dwell time, so one enters a quasiclassical regime where RMT no longer
holds. This quasiclassical regime has not yet been explored
experimentally. The difficulty is that $\tau_E$ increases so slowly with
system size that the averaging effects of inelastic scattering take over
before the effect of a finite Ehrenfest time can be seen. In a computer
simulation inelastic scattering can be excluded from the model by
construction, so this seems a promising alternative to investigate the
crossover from universal quantum fluctuations to nonuniversal quasiclassical
fluctuations.  Contrary to what one would expect from the disordered metal
\cite{Gal91}, where quasiclassical fluctuations are much smaller than the
quantum value, we find that the breakdown of universality in the chaotic
system is associated with an {\em{enhancement}} of the sample-to-sample fluctuations.

The quantity on which we choose to focus is the excitation gap $\varepsilon_0$
of a chaotic system which is weakly coupled to a superconductor. We have two
reasons for this choice: Firstly, there exists a model (the Andreev kicked
rotator) which permits a computer simulation for systems large enough that
$\tau_E\agt\tau_D$. So far, such simulations,
have confirmed the microscopic theory of Ref.~\cite{Vav03} for the average gap
$\langle \varepsilon_0 \rangle$ \cite{Jac03}. 
Secondly, there exists a quasiclassical
theory for the effect of a finite Ehrenfest time on the excitation gap
and its fluctuations \cite{Sil03}. This allows us to achieve both a numerical 
and an analytical understanding of the mesoscopic fluctuations when RMT breaks down.

We summarize what is known from RMT for the gap fluctuations \cite{Vav01}. In
RMT the gap distribution $P(\varepsilon_0)$ is a universal function of the
rescaled energy $(\varepsilon_0-E_g)/\Delta_g$, where $E_g=0.6\, E_T$ is the
mean-field energy gap and $\Delta_g=0.068\,g^{1/3}\,\delta$ determines the
mean level spacing just above the gap. The distribution function has mean
$\langle\varepsilon_0\rangle=E_g+1.21\,\Delta_g$ and standard deviation
$\left(\langle\varepsilon_0^2\rangle-\langle\varepsilon_0\rangle^2\right)^{1/2}
\equiv\delta\varepsilon_{\rm{RMT}}$ given by
\begin{equation}
\label{STDRMT}
\delta\varepsilon_{\rm{RMT}}=1.27\Delta_g=1.09\,E_T/g^{2/3}. 
\end{equation}
The RMT predictions for $P(\varepsilon_0)$, in the regime  $\tau_E\ll\tau_D$,
were confirmed numerically in Ref. \cite{Jac03} using the Andreev kicked 
rotator.

\begin{figure}
\includegraphics[width=8cm]{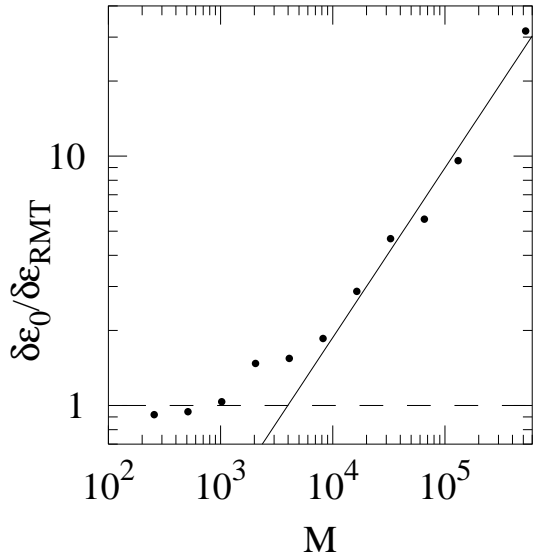}
\caption{Root-mean-square value $\delta\varepsilon_0$ of the gap divided by
  the RMT prediction $\delta\varepsilon_{\rm{RMT}}$, as a function of the
  system size $M$ for dwell time $M/N=5$ and kicking strength $K=14$. The data
  points result from the numerical simulation of the Andreev kicked
  rotator. The solid line has slope $2/3$, indicating that
  $\delta\varepsilon_0$ depends only on $M/N$ and not on $M$ or $N$ separately
  in the large-$M$ regime.}
\label{fluc}
\end{figure}

\begin{figure*}
\includegraphics[width=12cm]{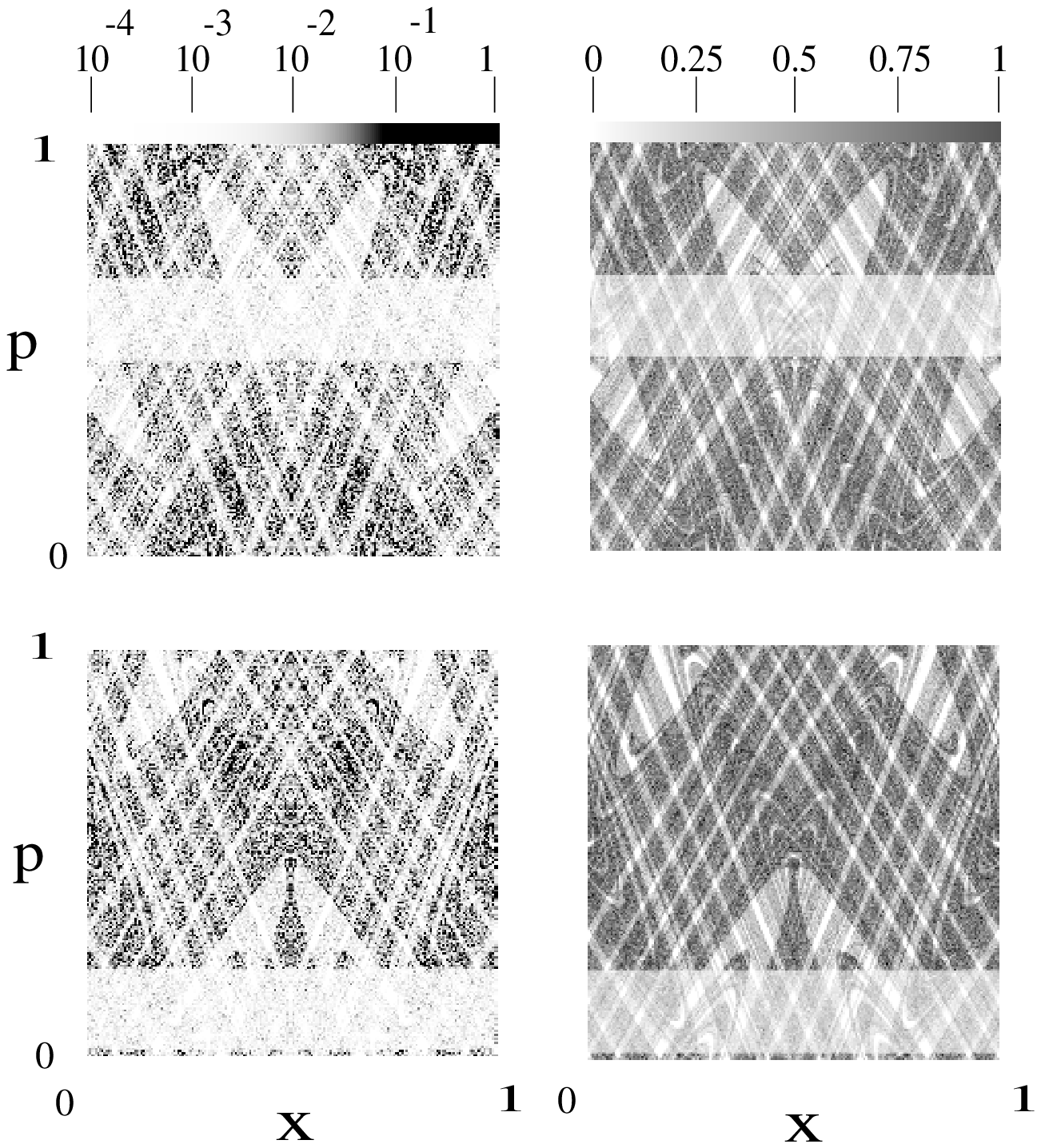}
\caption{Left panels: Husimi function (\ref{husimi}) for the electron
  component of the ground-state wavefunction $\psi_0$ of the Andreev kicked
  rotator, for two different positions of the lead. The parameters are
  $M=131072$, $\tau_D=M/N=5$, $K=14$. The calculated values are scaled by a
  factor $0.019$ ($0.017$) in the top (bottom) panel, so that they cover the
  range $(0,1)$, indicated by the gray scale at the top. Right panels: The
  corresponding classical density plots of all trajectories which have a time
  $t>7$ between Andreev reflections. The calculated values are rescaled by a
  factor $0.30$ ($0.32$) in the top (bottom) panel.}
\label{figure1}
\end{figure*}

We will use the same model, this time focusing on the gap fluctuations
$\delta\varepsilon_0$ in the regime $\tau_E\gtrsim\tau_D$. The Andreev kicked
rotator provides a stroboscopic description (period $\tau_0$) of the dynamics
in a normal region of phase space (area $M\hbar_{\rm{eff}}$) coupled to a
superconductor in a much smaller region (area $N\hbar_{\rm{eff}}$, $1\ll N\ll
M$). We refer to this coupling as a ``lead''. The effective Planck constant is
$\hbar_{\rm{eff}}=1/M$. The mean dwell time in the normal region (before
entering the lead) is $\tau_D=M/N$ and the corresponding Thouless energy is
$E_T=N/2M$. We have set $\tau_0$ and $\hbar$ equal to $1$. The dimensionless
conductance of the lead is $g=N$. The product $\delta=4\pi E_T/g=2\pi/M$ is
the mean spacing of the quasi-energies $\varepsilon_m$ of the normal region
without the coupling to the superconductor. The phase factors
$e^{i\varepsilon_m}$ ($m=1,2,..,M$) are the eigenvalues of the Floquet
operator $F$, which is the unitary matrix that describes the dynamics in the
normal region. In the model of the kicked rotator the matrix elements of $F$
in momentum representation are given by \cite{Izr90}
\begin{subequations}\label{floquet}
\begin{eqnarray}
F_{nm}&=&e^{-(i\pi/2M)(n^2+m^2)}(UQU^{\dag})_{nm},\\
U_{nm}&=&M^{-1/2}e^{(2\pi i/M)nm},\\
Q_{nm}&=&\delta_{nm}e^{-(iMK/2\pi)\cos{(2\pi n/M)}}.
\end{eqnarray}
\end{subequations}

The coupling to the superconductor doubles the dimension of the Floquet
operator, to accomodate both electron and hole dynamics. The scattering from
electron to hole, known as Andreev reflection, is described by the matrix
\begin{eqnarray}\label{projector}
{\cal{P}}^{1/2}=\left(
\begin{array}{cc}
1-(1-\frac{1}{2}\sqrt{2})P^TP & -i\frac{1}{2}\sqrt{2}P^TP\\
-i\frac{1}{2}\sqrt{2}P^TP & 1-(1-\frac{1}{2}\sqrt{2})P^TP
\end{array}
\right),
\end{eqnarray}
with the projection operator
\begin{eqnarray}\label{lead}
 \left(P^{T}P\right)_{nm}=\delta_{nm}\times\left\{\begin{array}{ll}
1& \mbox{if $L\le n\le L+N-1$,}\\
0& \mbox{otherwise.}
\end{array}\right.
\end{eqnarray}
Since we work in momentum representation, the lead defined by Eq.\
(\ref{lead}) is a strip in phase space of width $N$ parallel to the coordinate
axis. One could alternatively consider a lead parallel to the momentum axis,
if one would work in coordinate representation. We do not expect any
significant differences between the two alternatives.
Putting all this together we arrive at the Floquet operator of the Andreev
kicked rotator \cite{Jac03},
\begin{equation}\label{kandreev}
{\cal{F}}={\cal{P}}^{1/2}\left(
\begin{array}{ll}
F&0\\
0&F^*
\end{array}
\right){\cal{P}}^{1/2}.
\end{equation}
The matrix ${\cal{F}}$ can be diagonalized efficiently using the Lanczos
technique in combination with the Fast-Fourier-Transform algorithm
\cite{Ket99}. This makes it possible to calculate the quasi-energies
$\varepsilon_m$ and eigenfunctions $\psi_m$ for systems of sizes up to
$M=5\cdot 10^5$. The gap value $ \varepsilon_0$ is given by the eigenphase 
of ${\cal F}$ closest to zero.

The Floquet operator (\ref{kandreev}) provides a stroboscopic description 
of the
electron and hole dynamics, which is believed to be equivalent to the
true Hamiltonian dynamics on long time scales $t\gg \tau_{0}$. The
support for this comes from two sides: (i)~In the absence of 
superconductivity, and for varying parameters $K$ and
$\hbar_{\rm eff}$, the 1-D kicked rotator
correctly reproduces properties of localized \cite{Fish82},
diffusive \cite{Alt96}, and even ballistic \cite{Tajic03}
quasiparticles in disordered media.
(ii)~In the presence of superconductivity, the kicked Andreev rotator,
and extensions thereof, adequately describe quantum dots in
contact with a superconductor~\cite{Jac03}, and give a proper description
of quasiparticles in dirty d-wave superconductors \cite{Ada03}.
\begin{figure}
\includegraphics[width=8cm]{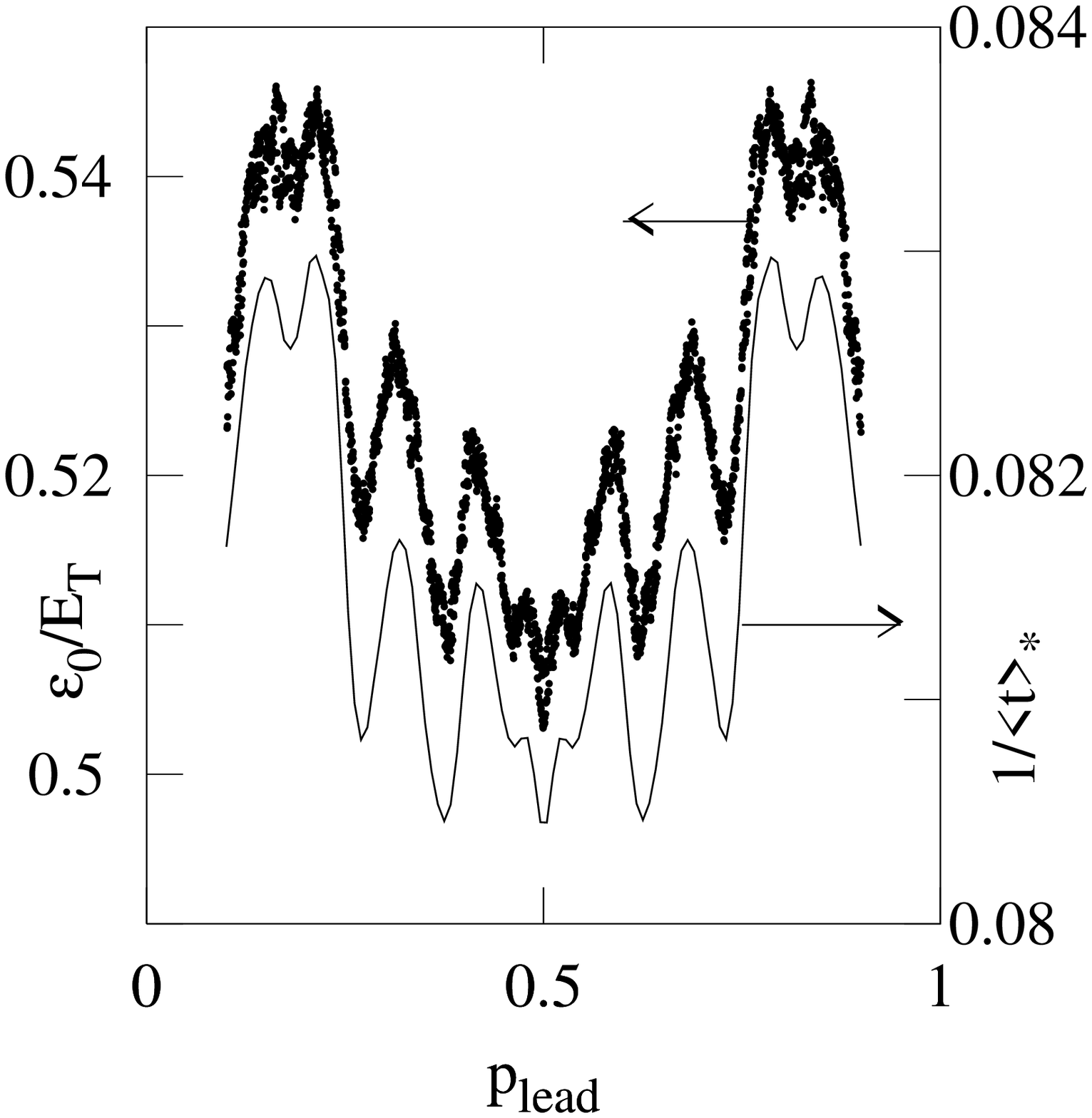}
\caption{The data points (left axis) are the quantum mechanical gap values
  $\varepsilon_0$ of the Andreev kicked rotator as a function of the position
  $p_{\rm{lead}}$ of the lead, for parameter values $M=131072$,
  $\tau_D=M/N=5$, $K=14$. The solid line (right axis) is the reciprocal of the
  mean dwell time $\langle t\rangle_*=\int_{t^*}^\infty
  tP(t)dt/\int_{t^*}^\infty P(t)dt$ of classical trajectories longer 
  than $t^*=7$.}
\label{Egap}
\end{figure}
Since we will be giving a classical interpretation of our results, we also
describe the classical map corresponding to the Andreev kicked rotator. The
map relates the dimensionless coordinate $x_n\in(0,1)$ and momentum
$p_n\in(0,1)$ at time $(n+1)\tau_0$ to the values at time $n\tau_0$:
\begin{subequations}\label{classical}
\begin{eqnarray}
p_{n+1}&=&p_n\pm \left(K/2\pi\right)\sin[2\pi(x_n\pm\frac{p_n}{2})],\\
x_{n+1}&=&x_{n}\pm \frac{p_n}{2}\pm \frac{p_{n+1}}{2}.
\end{eqnarray}
\end{subequations}
The upper and lower sign correspond to electron and hole dynamics,
respectively. Periodic boundary conditions hold both for $x$ and $p$. The
quasiparticle reaches the superconductor if  $|{p_{n+1}-p_{\rm{lead}}}|<
N/2M$, where $p_{\rm{lead}}$ is the center of the lead. At the next iteration
the electron is converted into a hole and vice versa.

We study a system with kicking strength $K=14$ (fully chaotic, Lyapunov
exponent $\lambda=1.95$) and vary the level spacing $\delta=2\pi/M$ at fixed
dwell time $\tau_D=M/N=5$.
Sample-to-sample fluctuations are generated by varying the position
$p_{\rm{lead}}$ of the lead over some $400$ locations. The resulting
$M$-dependence of $\delta\varepsilon_0$ is plotted in Fig.\ \ref{fluc} on a
double logarithmic scale. We have divided the value $\delta\varepsilon_0$
resulting from the simulation by the RMT prediction
$\delta\varepsilon_{\rm{RMT}}$ from Eq.\ (\ref{STDRMT}). The numerical data
follows this prediction for $M\alt 10^3$, but for larger $M$ the fluctuations
are bigger than predicted by RMT. For $M\agt10^4$ the ratio
$\delta\varepsilon_0/\delta\varepsilon_{\rm{RMT}}$ grows as $M^{2/3}$ (solid
line). Since $\delta\varepsilon_{\rm{RMT}}\propto M^{-2/3}$, this means that
$\delta\varepsilon_0$ is independent of the level spacing $\delta=2\pi /M$ at
fixed dwell time $\tau_D=M/N$. This suggests a quasiclassical explanation.

To relate the fluctuations of $\varepsilon_0$ to the classical dynamics, we
first examine the corresponding wavefunction $\psi_0$. In the RMT regime the
wavefunctions are random and show no features of the classical
trajectories. In the quasiclassical regime $\tau_E\agt\tau_D$ we expect to see
some classical features. Phase space portraits of the electron components
$\psi_m^e$ of the wavefunctions are given by the Husimi function
\begin{equation}
\label{husimi}
{\cal{H}}(n_x,n_p)=|\langle\psi_m^e|n_x,n_p\rangle|^2.
\end{equation}
The state $|n_x,n_p\rangle$ is a Gaussian wave packet centered at $x=n_x/M$,
$p=n_p/M$. In momentum representation it reads 
\begin{equation}
\langle n|n_x,n_p\rangle\propto e^{-\pi(n-n_p)^2/M}e^{2\pi i n_x n/M}.
\end{equation}

In Fig.\ \ref{figure1}, left panels, the Husimi function of $\psi_0$ is shown
for two lead positions. Shown is a logarithmic gray scale density plot of the
Husimi function, with light (dark) areas corresponding to low (high)
density. The lead is visible as a light strip parallel to the $x$-axis. It is
clear that these wavefunctions are not random. We expect that the structure
that one sees corresponds to long classical trajectories, since the
wavefunctions are for the lowest quasi-energy. To test this expectation, we
show in the right panels (on a linear gray scale) the corresponding classical
density plots for all trajectories with dwell time $t>t^*$. A total of $3\cdot
10^5$ initial conditions $(x_0,p_0)$ for these trajectories are chosen
uniformly in the lead. Each new iteration of the map (\ref{classical}) gives a
point $(x_n,p_n)$ in phase space, which is kept if the time of return to the
lead is greater than $t^*$. We take $t^*=7$, somewhat larger than the
Ehrenfest time $\tau_E=\lambda^{-1}\ln{(N^2/M)}=4.4$. The plot is not
particularly sensitive to the value of $t^*$, as long as $t^*>\tau_E$. There
is a clear correspondence between the quantum mechanical Husimi function and
the classical density plot. We conclude that the wavefunction of the lowest
excitation covers predominantly that part of phase space where the longest
dwell times occur.

To make this more quantitative we show in Fig.\ \ref{Egap} the gap value from
the quantum  simulations as  a function of the lead position. The solid curve
results from a classical calculation of the mean dwell time of those
trajectories with $t>t^*$, for the same value $t^*=7$ used in Fig.\
\ref{figure1}. More precisely, it is a plot of 
\begin{equation}
\frac{1}{\langle t \rangle_*}=
\frac{\int_{t^*}^{\infty}P(t)dt}{\int_{t^*}^{\infty}tP(t)dt},
\end{equation}
with $P(t)$ the classical dwell time distribution. We see that the
sample-to-sample fluctuations in the gap $\varepsilon_0$ correlate very well
with the fluctuations in the sample-to-sample mean dwell time of long
trajectories. Again, the correlation is not sensitive to the choice
$t^*>\tau_E$. Such a correlation is in accord with recent theoretical work
\cite{Sil03}, in which an effective RMT description is expected to hold for
the part of phase space with dwell times greater than the Ehrenfest time. But
we should emphasize that the agreement is only qualitative. In particular, the
relation $\varepsilon_0\approx 1.5/\langle t\rangle_*-0.07$ that we infer from
Fig.\ 3 is different from the relation $\varepsilon_0=0.3/\langle t \rangle
_*$ that would be expected from RMT. While the theory of Ref.~\cite{Vav03}
has been found to be in good agreement with the average gap value
$\langle \varepsilon_0 \rangle$ \cite{Jac03}, it is not clear how it
compares to the data of Fig.\ 3.

In conclusion, we have investigated the transition from quantum mechanical to
quasiclassical gap fluctuations in the superconductor proximity effect. The
transition is accompanied by a loss of universality and a substantial
enhancement of the fluctuations. Our numerical data provides qualitative
support for an effective random-matrix theory in a reduced part of phase space
\cite{Sil03}, as is witnessed by the precise correlation which we have found
between the value of the gap and the dwell time of long classical trajectories
(see Fig.\ 3). 
It would be of interest to investigate to what extent
quasiclassical fluctuations of the conductance in a ballistic chaotic system
are similar or different from those of the superconducting gap studied here.

We have benefitted from discussions with J. Tworzyd{\l}o. This work was
supported by the Dutch Science Foundation NWO/FOM and the Swiss National
Science Foundation.

\end{document}